\documentclass[12pt]{article}
\usepackage{amsmath,amsfonts,amssymb,bbm}
\usepackage{graphics}

\usepackage[latin1]{inputenc}
\usepackage[pdftex]{graphicx}

\usepackage{fullpage}
\usepackage{color}
\usepackage[american]{babel}

\makeatletter \@addtoreset{equation}{section}

\makeatletter\renewcommand\section{\@startsection {section}{1}{\z@}%
                                   {-3.5ex \@plus -1ex \@minus -.2ex}
                                   {2.3ex \@plus.2ex}%
                                   {\normalfont\large\bfseries}}
\renewcommand\subsection{\@startsection{subsection}{2}{\z@}%
                                     {-3.25ex\@plus -1ex \@minus -.2ex}%
                                     {1.5ex \@plus .2ex}%
                                     {\normalfont\bfseries}}

\renewcommand{\baselinestretch}{1.2}

\newcommand{\be}{\begin{equation}}
\newcommand{\ee}{\end{equation}}
\newcommand{\bea}{\begin{eqnarray}}
\newcommand{\eea}{\end{eqnarray}}
\newcommand{\bmat}{\begin{bmatrix}}
\newcommand{\emat}{\end{bmatrix}}

\newcommand{\bbibitem}[1]{\bibitem{#1}\marginpar{#1}}

\def\Label#1{\label{#1}%
  \smash{\hbox to0pt{\raise1ex\hbox{\tiny[#1]}\hss}}}
\def\noLabels{\let\Label=\label}
\def\nobbibitem{\let\bbibitem=\bibitem}

\def\CL{{\cal L}}

\def\CW{{\cal W}}

\newcommand{\RR}{\mathbb{R}}
\newcommand{\ZZ}{\mathbb{Z}}

\newcommand{\R}{\mathbb{R}}

\begin{document}

\begin{titlepage}

\vfil\

\begin{center}

{\Large{\bf Black Holes and Singularity Resolution in Higher Spin Gravity}}

\vspace{3mm}

 Alejandra Castro\footnote{e-mail: acastro@physics.mcgill.ca}$^{a}$,
Eliot Hijano\footnote{email: eliot.hijano@mail.mcgill.ca}$^{a}$,
 Arnaud Lepage-Jutier\footnote{e-mail: arnaud.lepage-jutier@mail.mcgill.ca}$^{a}$\\ \&
Alexander Maloney\footnote{e-mail: maloney@physics.mcgill.ca}$^{a}$
\\

\vspace{8mm}

\bigskip\medskip
\smallskip\centerline{$^a$ \it
McGill Physics Department, 3600 rue University, Montreal, QC H3A 2T8, Canada}
\medskip
\vfil

\end{center}
\setcounter{footnote}{0}
\begin{abstract}
\noindent

We investigate higher spin theories of gravity in three dimensions based on the
gauge group $SL(N, \RR)\times SL(N, \RR)$.  In these theories the
usual diffeomorphism symmetry is enhanced to include higher spin
gauge transformations under which traditional geometric notions of
curvature and causality are no longer invariant.  This implies, for
example, that apparently singular geometries can be rendered smooth
by a gauge transformation, much like the resolution of orbifold
singularities in string theory.  The classical solutions, including
the recently constructed higher spin black hole, are characterized
by their holonomies around the non-contractible cycles of
space-time. The black hole solutions are shown to be gauge equivalent to a BTZ
black hole which is charged under a set of $U(1)$ Chern-Simons
fields.  Nevertheless, depending on the choice of embedding of the gravitational gauge group, the space-time geometry may 
be non-trivial.
We study in detail the $N=3$ example, where this observation allows us to find a gauge where the black hole 
geometry takes a simple form and the thermodynamic properties can be studied.

\end{abstract}
\vspace{0.5in}

\end{titlepage}
\renewcommand{\baselinestretch}{1.1}  

\newpage
\tableofcontents

\section{Introduction}

Three dimensional general relativity has proven a useful testing ground for our ideas about classical and quantum gravity.  The theory is locally trivial, yet exhibits a range of interesting phenomena from black holes to holography, allowing one to address deep questions about quantum gravity in an exactly solvable setting.  Recent attention has focused on extensions of pure general relativity to include additional massless higher spin degrees of freedom.  In these theories the standard general coordinate invariance of general relativity is extended to include higher spin versions of the diffeomorphism group.  These symmetries mix the metric and the higher spin degrees of freedom, requiring us to revisit basic notions of geometry and causality.

Theories with interacting massless higher spin degrees of freedom
are notoriously complicated.  Four dimensional theories have been
constructed by Vasiliev, but have proven difficult to study
explicitly. For a review see \cite{Vasiliev:2001ur,Bekaert:2005vh}. In three dimensions, however, the construction of such
theories is nearly trivial.  General relativity with a negative
cosmological constant is classically equivalent to a Chern-Simons
theory with $SO(2,2)$ gauge group.  One can then simply extend the
gauge group to obtain a more complicated theory.  In terms of the
usual second order formulation of the theory, the low energy degrees
of freedom will include a massless graviton as well as linearized
massless higher spin fields in Anti-de Sitter Space.  The field
content of the theory will depend on the choice of gauge group as
well as on how the gravitational $SO(2,2)$ is embedded into this
larger gauge group.  In this paper we consider the case where the
gauge group is $SL(N,\RR)\times SL(N,\RR)$, and focus on two
possible embeddings of the gravitational $SO(2,2)$, known as the
principal and diagonal embedding.

Our interest in these theories is threefold.  First, they provide
interesting counterexamples to the Coleman-Mandula theorem, which
constrains possible extensions of the Lorentz group of interacting
theories. In the present case the three dimensional Anti-de Sitter
group is extended in a non-trivial way. Of course, these theories
merely violate the ``spirit" of the Coleman-Mandula theorem rather
than the technical proof.  For example, the resulting extended
symmetries are not generated by traditional Lie algebras and thus lie
outside the scope of the theories originally considered by Coleman
and Mandula \cite{Coleman:1967ad}.

Second, these theories may provide examples of exactly solvable theories of pure gravity with a semiclassical limit.  Despite notable efforts, there are only a handful of cases where known CFTs can be argued to be dual to purely metric theories of gravity (see e.g. \cite{Witten:2007kt,Chamon:2011xk}). These theories have central charges of order one and therefore describe gravity in a highly quantum mechanical regime.  However, for higher spin theories it may be possible to identify theories with large central charge that appear to be dual to pure (higher spin) theories of gravity.  For example, while the usual Virasoro minimal models exist only for small values of the central charge, higher spin generalizations of these minimal models exist for large values of the central charge.  This has led to a conjectured set of dualities between specific higher spin theories of the sort discussed in this paper and certain higher spin minimal models \cite{Gaberdiel:2010pz}.  This is the three dimensional version of the conjecture by Klebanov and Polyakov which relates higher spin Vasiliev theories to the critical $O(N)$ model \cite{Klebanov:2002ja} (see also \cite{Sezgin:2002rt,Giombi:2009wh,Giombi:2010vg}).

Finally, and perhaps most importantly, these theories share many features with string theory.  In the tensionless limit the higher spin perturbative string states become massless and should therefore act as gauge fields for higher spin symmetries.  Thus our higher spin theories can be regarded as models of the tensionless limit of string field theory in AdS backgrounds.  This so-called ``unbroken phase" of string theory should exhibit many interesting features and should shed light on various aspects of the AdS/CFT correspondence at weak coupling.  Indeed, our higher spin theories of gravity share one important qualitative feature of string theory, in that traditional notions of geometry must be modified.  The standard diffeomorphism invariant quantities, such as curvature and causal structure, are no longer invariant under the higher spin gauge symmetries.  Physical observables will not depend on certain geometric properties of a given solution, a common feature in string dualities.

It is therefore natural to ask whether one of the most important features of
string theory -- singularity resolution -- can be realized in the
context of higher spin gravity.  String propagation is unitary on a
certain apparently singular manifolds; this can be viewed as a
consequence of an enhanced symmetry underlying string worldsheet
theory, as in \cite{Dixon:1985jw, Dixon:1986jc, Witten:1991yr}.  We
will see that in a sense the same is true of higher spin theories,
as the higher spin gauge transformations can render non-singular an apparently
singular geometry.  Thus some of the usual pathologies
associated with curvature singularities are no longer present.
However we will see that this singularity resolution comes at a
price, in that other higher spin fields are turned on and the causal
structure of space-time is altered.

We subsequently turn to the study of black holes which carry charge under
the higher spin gauge fields.  In the
Chern-Simons formulation, such solutions are characterized completely
by their holonomies around a single non-contractible cycle.  These solutions
can be most conveniently understood by noting that the Chern-Simons
gauge theory of interest contains as a consistent truncation an $SL(2,\RR)\times SL(2,\RR)$ gauge
theory
coupled to a set of $U(1)$ Chern-Simons gauge theories.  This theory is classically
equivalent to Einstein gravity coupled to $U(1)$ gauge theories, so the charged BTZ black hole
solutions of this theory can be easily uplifted to solutions of the full $SL(N,\RR)\times SL(N,\RR)$ gauge
theory. All known black hole solutions of the theory, including those described recently by
\cite{Didenko:2006zd, Gutperle:2011kf, Ammon:2011nk, Kraus:2011ds}, can be constructed in this manner.\footnote{See also \cite{Didenko:2009td,Iazeolla:2011cb} for a discussion of higher spin black holes in four dimensions.} The geometry of such a solution may still differ from that of the BTZ black hole, however,
depending on the choice of embedding.  Nevertheless, the holonomy characterization of the solutions allows us to construct gauges where the geometries take a relatively simple form and their thermodynamics properties can be studied.  In the diagonal embedding, the solutions and thermodynamics are essentially identical to those of a charged BTZ black hole.  In the principal embedding, although the connection is the same, the gravitational interpretation is different.

We begin with a review of higher spin gauge theory in section 2, emphasizing the qualitative differences between the choices of gravitational embedding.  We discuss the holonomy classification of solutions in section 3.  In section 4 we demonstrate singularity resolution, before discussing black holes in section 5.

\section{Higher Spin Gravity in Three Dimensions}\label{sec:HS}

We now summarize the relevant features of higher spin theories in three dimensional gravity with a negative cosmological constant.  For further details and definitions see e.g.  \cite{Aragone:1983sz,Blencowe:1988gj,Bergshoeff:1989ns,Campoleoni:2010zq}.


Three dimensional general relativity has no local degrees of freedom and at the classical level is equivalent to a Chern-Simons gauge theory \cite{Achucarro:1987vz,Witten:1988hc,Achucarro:1989gm}. The gauge group  depends on the sign of the cosmological constant; for negative cosmological constant it is $SO(2,2)= (SL(2,\RR)\times SL(2,\RR))/\ZZ_2$.\footnote{For the time being we will omit the $\ZZ_2$ factor and simply take the gauge group to be $Spin(2,2)=SL(2,\RR)\times SL(2,\RR)$.  This will not make any difference until we discuss spinor holonomies in section 5, when we will return to this global issue in more detail.}  Let us first recall this construction.
We begin by combining the dreibein $e_\mu^{~a}$ and the spin connection $\omega_{\mu}^{~a}$ into a pair of $SL(2,\RR)$ gauge fields
\bea\label{sec2:aa}
A_{(2)}=J_a\left(\omega_{\mu}^{~a}+{1\over \ell}e_\mu^{~a}\right)dx^\mu~,\quad
\bar A_{(2)}=J_a\left(\omega_{\mu}^{~a}-{1\over \ell}e_\mu^{~a}\right)dx^\mu~.
\eea
Here $J_a$ are the generators of $sl(2,\RR)$.  The Einstein-Hilbert action is
given by
\be\label{sec2:ba}
I_{EH}=\left(I_{CS}[A_{(2)}]-I_{CS}[\bar A_{(2)}]\right)~.
\ee
The Chern-Simons action at level $k$ is
\be\label{sec2:ad}
I_{CS}[A]={k \over 4 \pi} \int {\rm tr}(A\wedge dA+{2\over 3}A\wedge A\wedge A)~,
\ee
where ${\rm tr}$ is the symmetric bilinear form on $SL(2,\RR)$.
The level $k$ of the theory is determined by the AdS$_3$ radius $\ell$ and Newton's constant  $G_3$
\be
 k={\ell\over 4G_3}~.
\ee
The Chern-Simons equations of motion imply that the connections $A_{(2)}$ and $\bar A_{(2)}$ are flat.  This is equivalent to Einstein's equation with a negative cosmological constant.

We wish to generalize this theory to include more interesting degrees of freedom.  A simple way to do so is to extend the gauge group in some way.
The simplest example, which is the subject of this paper, is found by taking the gauge group to be $G= SL(N, \RR)\times SL(N,\RR)$ \cite{Campoleoni:2010zq}.  This will describe AdS${}_3$ gravity coupled to additional higher spin degrees of freedom.
Of course, the theory still has no local propagating degrees of freedom.

To obtain the low energy field content we must linearize the equations of motion around empty AdS.  This requires us to choose how the gravitational $SL(2,\RR)\times SL(2,\RR)$ subgroup is embedded into the full  $SL(N,\RR)\times SL(N,\RR)$ gauge group.  In fact, the $N>2$ Chern-Simons theory can be interpreted as a variety of different higher spin theories in AdS${}_3$ depending on how we choose this embedding.  In this paper we will focus on two possible choices, known as the principal embedding and the diagonal embedding.  These embeddings are not related by conjugation in $SL(N,\RR)$, so they define two physically distinct extensions of general relativity in AdS${}_3$.

We start by describing the principal embedding.  In this case the linearized fields describe a massless spin-2 field (the graviton) coupled to a tower of massless symmetric tensor fields with spin running from $3$ up to $N$.   This case is perhaps the closest three dimensional analogue of the four dimensional Vasiliev theories.  In particular, one can take the infinite dimensional extension of $SL(2,\RR)$ -- denoted $hs(1,1)$ -- which will describe a infinite tower of spins just as in the Fradkin-Vasiliev theory \cite{Blencowe:1988gj,Bergshoeff:1989ns}.  

To understand the structure of the principal embedding, let us focus on the case $N=3$,  describing gravity coupled to a spin-3 field.  The simplest way to understand the theory is to introduce a basis of generators  $J_a$ and $T_{ab}$ for the $sl(3, \RR)$ Lie algebra, which obey
\bea\label{algebra}
[J_a,J_b]&=&\epsilon_{abc}J^c~,\cr [J_a,T_{bc}]&=&\epsilon^m_{~~a(b}T_{c)m}~,\cr
 [T_{ab},T_{cd}]&=&-(\eta_{a(c}\epsilon_{d)bm}+\eta_{b(c}\epsilon_{d)am})J^{m}~.
\eea
The first line states that the $J^a$ obey the $sl(2,\RR)$ algebra.  The second line states that the symmetric, traceless tensors $T_{ab}$ transform in the spin-2 representation of $sl(2,\RR)$.
In addition to the dreibein and spin-connection given above, we can introduce the tensor valued one-forms %
\bea\label{sec2:ax}
\quad e_\mu^{~a b}~, \quad\omega_\mu^{~ab}~.
\eea
If we expand the $sl(3,\RR)$ connections as
\bea\label{sec2:axa}
A&=& A_{(2)}+T_{ab}\left(\omega_\mu^{~ab}+{1\over \ell}e_\mu^{~ab}\right)dx^\mu~,\cr
\bar A&=& \bar A_{(2)}+T_{ab}\left(\omega_\mu^{~ab}- {1\over \ell}e_\mu^{~ab}\right)dx^\mu~,
\eea
then $e_{\mu}^{ab}$ and $\omega_\mu^{ab}$ can be regarded as the frame fields and connection associated to the spin-3 gauge symmetry generated by $T_{ab}$.%

The action of the theory is
\be\label{sec2:ba}
I_{N}=I_{CS}[A]-I_{CS}[\bar A]~,\quad  k={\ell\over 4G_3\epsilon}~,
\ee
where $I_{CS}$ is defined using a symmetric bilinear form ${\rm tr}$ on the $sl(3, \RR)$ algebra.
The factor of $\epsilon$ arises because when we  restrict to the gravitational $SL(2,\RR)\times SL(2,\RR)$ subgroup the  trace on $sl(3,\RR)$ will reduce to the standard metric on $sl(2,\RR)$ only up to an overall normalization factor.  To reproduce  (\ref{sec2:ba}) we must therefore define $\epsilon$ to be ${\rm tr}(J_a J_b)=\epsilon\eta_{ab}/2$.
For the standard metric on $sl(3,\RR)$, presented in appendix \ref{app:conv}, we have $\epsilon=4$.

Given these gauge fields it is straightforward to construct the space-time metric and spin-3 field
\be\label{defgab}
g_{\mu\nu}={1\over 2}{\rm tr}(e_{(\mu} e_{\nu)})~,\quad \psi_{\mu\nu\alpha}={1\over 9}{\rm tr}(e_{(\mu}e_\nu e_{\alpha)})~.
\ee
Here $e_\mu=e_\mu^{~a}J_a+e_\mu^{~ab}T_{ab}$. One can check that the linearized fluctuations of the gauge field $e_\mu^{~ab}$ around the AdS background satisfy the equation of motion of a higher spin field as defined by Fronsdal \cite{Fronsdal:1978rb}.
At the linearized level this spin-3 field will possess gauge symmetries generated by the $T_{ab}$.
Therefore the action \eqref{sec2:ba} describes the dynamics of a gravitational theory coupled to a spin-3 field.

At the non-linear level this action will contain complicated interactions between gravity and the higher spin field.  The gauge transformations will mix the metric and the spin-3 field in a complicated way.  This theory should therefore be regarded as a more symmetric generalization of general relativity, where the diffeomorphism symmetries (generated by spin-1 generators $J_a$ in the Chern-Simons formulation) are extended to include higher spin symmetries.  This leads to many interesting effects.  For example, standard diffeomorphism invariant notions -- such as causality and curvature invariants built out of the metric -- are no longer gauge invariant and hence are not necessarily meaningful physical quantities.  This will lead to many of the interesting effects considered in this paper.

The other embedding which will be considered in this paper is the diagonal embedding, where $SL(2,\RR)$ is simply taken to be a block diagonal $2 \times 2$ matrix in $SL(N,\RR)$.  In this case the remaining generators will transform in $2(N-2)$ spin-1/2 representations of $SL(2,\RR)$ and $(N-2)^2$ spin-0 representations.  The linearized fluctuations around the empty AdS${}_3$ solution can be computed as above.  Each of these spin-$0$ representations will give a standard spin-1 gauge field, while each spin-$1/2 $ generator leads to a bosonic spin-3/2 gauge field.\footnote{This is possible because there is no spin-statistics theorem for massless fields in three dimensions \cite{Binegar:1981gv,Deser:1991mw}.}  As before, at the non-linear level these gauge fields will interact with each other and with the gravitational field in a complicated manner.

With the diagonal embedding, the theory contains an important truncation, where the connection is taken to lie in the block diagonal  $SL(2,\RR)\times U(1)^{N-2}$.  This truncation is just AdS${}_3$ gravity coupled to $U(1)^{2(N-2)}$ Chern-Simons gauge fields.  The solutions of this theory are locally AdS${}_3$ metrics equipped with $2(N-2)$ flat $U(1)$ connections.  This includes black holes which are charged under the $U(1)$ Chern-Simons gauge fields.  These solutions immediately lift to solutions of the full higher spin theory in the diagonal embedding.

The metric in the diagonal embedding is constructed as follows.  We use a basis of generators $\{\hat J_a, U_i, S_n^{\pm}\}$ where $\hat J_a$ is the $sl(2,\RR)$ subgroup,  $U_i$ are those elements that transform in the spin-0 representation and each pair $S_n^{\pm}$ transform in the spin-1/2 representation. We write the connection as
\bea
A= A_{(2)}+ \chi^i U_i + \Psi^n  S_n~,\cr
\bar A = \bar A_{(2)}+ \bar \chi^i U_i + \bar \Psi^n S_n~,
\eea
 where $A_{(2)}$ and $\bar A_{(2)}$ are given by \eqref{sec2:aa} with generators $\hat J_a$. Here ${\chi^i, \bar \chi^i}$ are one forms and $(\Psi^n,\bar\Psi^n)$ are two components spinors
 \be
 \Psi^n=\left(\begin{array}{c} \Psi^{+,n} \\  \Psi^{-,n}\end{array}\right) ~, \quad \Psi^nS_n= \Psi^{+,n} S^{+}_n +\Psi^{-,n} S^{-}_n~.
 \ee

 In contrast with the principal embedding, the gauge fields $\chi^i$, $\Psi^n$ and barred counterparts naturally obey first order  equations of motion.  Thus it is not necessary to include an auxiliary field to accompany them. The metric is then simply given by
 \be\label{sec2:da}
 g_{\mu\nu}={1\over 2}{\rm Tr} (e_{(\mu} e_{\nu)})~,
 \ee
 with $e_\mu=e^a_\mu \hat J_a={1\over 2}(A_{(2)}-\bar A_{(2)})$.

In the following sections we will discuss $N=3$ and the interpretation of the solution in both the principal and diagonal embedding. Rather than using the $sl(3,\RR)$ generators $\{J_a, T_{ab}\}$ or  $\{\hat{J}_a, U_i, S_n\}$, we will find it convenient to use a basis of generators which accommodates both embeddings.  We will  use the fundamental representation of the matrices where we label the elements as   $L_a=\{L_{0},L_{\pm 1}\}$ and $W_m=\{W_0, W_{\pm1},W_{\pm 2}\}$, see appendix \ref{app:conv} for more details. The connections will be written as
\be
A=A^a L_a +A^{m} W_m ~,\quad \bar A=\bar A^a L_a +\bar A^{m} W_m ~.
\ee
In this basis the principal embedding generators are certain linear combinations of the $L_a$ and $W_m$.  Schematically,
\be
J_a=\{L_0,L_{\pm1}\}~,\quad T_{ab}=\{W_0,W_{\pm1},W_{\pm 2}\}~.
\ee
The precise relationship is given in appendix \ref{app:conv}.   Likewise, in the diagonal embedding the generators are linear
combinations%
\be
\hat J_a=\{{1\over 2}L_0, \pm{1\over 4}W_{\pm2}\}~,\quad U_1=W_0 ~,\quad S_1=\{L_{1},W_{-1}\}  ~,\quad S_2=\{L_{-1},W_{1}\}~.
\ee

\section{The Classical Solutions \& Holonomies}\label{sec:Hol}

The classical equations of motion of Chern-Simons theory state that the connection $A$ is flat
\be
dA + A \wedge A = 0~.
\ee
A flat connection is locally pure gauge, so can locally be written in the form
\be\label{gauge}
A = g^{-1} dg~.
\ee
where $g$ is a map from space-time into the gauge group.  This statement is not true globally, as the space-time may have non-trivial topology.

More precisely, if space-time has a non-contractible cycle $C$ then the holonomy of $A$ around $C$ is an element of the gauge group
\bea\label{Ha}
{\rm Hol}_C (A)={\cal P}\exp(\oint_C A )
\eea
This holonomy transforms by conjugation under a trivial gauge transformation.
If this holonomy is non-trivial then it is impossible to find a globally defined gauge transformation $g$ such that $A = g^{-1} dg$. If we were to try to do so, we would find that the function $g$ would not be a single valued function of space-time.  Instead, as we go around the cycle $C$ the gauge transformation $g$ will pick up a factor of the holonomy.

These holonomies described above are the only obstruction to writing the connection as pure gauge.
This means that flat connections are, up to an overall gauge transformation, uniquely specified by their holonomies around the non-contractible cycles of space-time.  In other words, classical solutions are uniquely labelled by maps from the fundamental group of space-time into the gauge group, modulo an overall conjugation by $G$.

In this paper we will focus on space-times where the connection has non-trivial holonomy around  a single non-contractible cycle.  In this case the solutions are uniquely labelled by the conjugacy classes of $G$, which represent the holonomy around this cycle.   For this reason it is worth describing the conjugacy classes of the gauge group $SL(N,\RR)$ in a bit more detail.  Any element $g$ of $SL(N,\RR)$ can, by conjugation, be put in real Jordan normal form.  The real Jordan form of $g$ is block diagonal, with one block for each eigenvalue $\lambda$ of $g$.  The simplest case is where  $\lambda$ is real and non-degenerate, in which case the block is simply the $1 \times 1$ matrix $\lambda$.  If $\lambda$ is real with degeneracy $n>1$ then the block may not be diagonalizable; if the number of eigenvectors is less than $n$, the block is the $n \times n$ matrix with $\lambda$ on the diagonal and ones just above the diagonal.  When $\lambda = r e^{i \theta}$ is complex, then situation is the same except that each $1 \times 1$ matrix $\lambda$ is replaced by the $2 \times 2 $ matrix
\be
\left({r \cos \theta ~ r \sin \theta \atop - r \sin\theta~r \cos\theta \\}\right)
\ee
and each $1$ is replaced by the $2 \times 2$ identity matrix.
From the point of view of Chern-Simons theory, this provides a complete classification of solutions to the equations of motion with one non-trivial cycle.

It is worth comparing this classification of solutions to the standard classification of solutions of Einstein gravity with a negative cosmological constant.  The solutions of the equations of motion with a single non-contractible cycle  are quotients of the AdS${}_3$ manifold by an element $(g_L, g_R)$ of $SL(2,\RR)\times SL(2,\RR)$ \cite{Maldacena:1998bw, Martinec:1998wm}.  The elements $(g_L, g_R)$ are precisely the holonomies described above in the case $N=2$.  When $(g_L, g_R)$ are diagonalizable with real eigenvalues the corresponding solution is the non-extremal BTZ black hole.  When $(g_L, g_R)$ have complex eigenvalues the corresponding solution describes a conical deficit in AdS.  In the degenerate case, where either matrix is (or both matrices are) non-diagonalizable, the corresponding solution is an extremal (or a massless) BTZ black hole.

In this framework it is  clear how to generalize the BTZ solution to form more complicated black hole solutions which carry higher spin charge.  If we are interested in solutions which are generalizations of the non-extremal BTZ black hole then we should choose a connection whose holonomy matrix has real, non-degenerate eigenvalues.\footnote{If the holonomies have non-degenerate eigenvalues then the conserved charges will all be independent parameters.  The solution will therefore be the higher spin version of a non-extremal black hole.}  In this case the solutions are uniquely labelled by the eigenvalues of the holonomy matrix.  Since there are two holonomy matrices, each of which lives in $SL(N,\RR)$,  there are $2(N-1)$ independent eigenvalues.  In the pure gravity $SL(2,\RR)$ case these will be the mass and angular momentum of the black hole.  

In the case of the $SL(3,\RR)$ theory we will have two independent eigenvalues for
each connection $A$ and $\bar A$. Therefore a black hole can carry
four independent charges; these black holes will be considered in more detail below.

Before doing so, let us be more specific about the connections under consideration. We start with a set of coordinates $(t, \phi, \rho)$ which label points in space-time.  The $\rho$ coordinate is a radial coordinate and the boundary will in general be given by $\rho \to \infty$.  The $\phi$ coordinate will be periodic with period $2\pi$ and will usually describe a non-contractible cycle.
Because $A$ has non-trivial holonomy it will not necessarily be single valued function of $\phi$; as one moves around a non-contractible cycle it will return to itself only up to a gauge transformation.
We will choose to work in a gauge where the radial component  is $A_\rho=b^{-1}\partial_\rho b$, where
$b$ is a function of $\rho$.
The function $b$ can be regarded as a gauge transformation, i.e. a single valued map from the manifold into the gauge group.
In general our solutions will take the form
\be\label{CSsol}
A= b^{-1}\, a\, b +b^{-1}\,db~,
\ee
and similarly for $\bar A$.
The flat connection $a$ will have non-trivial holonomy and be non-single valued.  The holonomy around the $\phi$-cycle is
\bea
{\rm Hol}_\phi (A)=b^{-1}\,\exp(\oint a_\phi\,d\phi)\,b~.
\eea
This holonomy will, up to gauge transformation, uniquely specify the solution.

Before proceeding, we note that the holonomy matrix itself might be difficult to evaluate for a given solution.  Instead, it will be convenient to describe the matrix by its characteristic polynomial. For example, any $3\times 3$ matrix $X$ can be decomposed as a linear combination of lower powers of the same matrix
\be
X^3=\Theta_0 \mathbbm{1} +\Theta_1X +\Theta_2 X^2~,
\ee
with $\Theta_i$ constant. Further, for $X\in sl(3,\R)$ one can check that $\Theta_2=0$ and
\be\label{inv}
\Theta_0= \det(X)~,\quad  \Theta_1={1\over 2}{\rm tr}(X^2) ~.
\ee
Therefore, if the eigenvalues are non-degenerate, the holonomy of a $sl(3,\RR)$ matrix is completely specified by $\Theta_0$ and $\Theta_1$.
%

\section{Singularity Resolution in Higher Spin Gravity}

Having formally described solutions in the Chern-Simons language, we now investigate their features from the metric point of view.
Although the solutions are in principle classified by their holonomies,  the geometrical interpretation of a given configuration at this stage is obscure. Indeed, the variables that define the geometric fields \eqref{defgab} and \eqref{sec2:da} are not gauge invariant so there may be multiple different geometric interpretations of a given classical solution.  We will now illustrate this fact using a  simple example.  We will consider solutions with fixed holonomy, and demonstrate that in some circumstances the corresponding metric can be either smooth or singular.  


We will consider the $SL(3,\R)$ case, though similar comments can be made for any value of $N$.
Consider connections in the pure gravitational sector of the principal embedding
\bea\label{aa}
A&= &(e^\rho L_1 -\CL e^{-\rho} L_{-1})dx^++L_0d\rho~,\cr
\bar A&=& -(e^\rho L_{-1} - \CL e^{-\rho} L_1)dx^--L_0d\rho~,
\eea
with $x^\pm =t\pm \phi$ and $\phi\sim\phi +2\pi$. The metric constructed from \eqref{aa} is
\be\label{metric}
ds^2=d\rho^2-\left(e^\rho -\CL e^{-\rho}\right)^2dt^2+\left(e^\rho +\CL e^{-\rho}\right)^2d\phi^2~,
\ee
and the spin-3 field is zero. Indeed, in the standard $SL(2,\R)$ formulation of Chern-Simons gravity the solution \eqref{aa} describes a solution with non-trivial holonomy in the $\phi$ direction.  For example, when $\CL\ge0$, this solution is the non-rotating BTZ black hole with mass $\CL$. When $-1/4<\CL<0$ the  metric has a conical singularity.  The value $\CL=-1/4$ corresponds to global AdS$_3$.   The invariants $\Theta_i$ that label ${\rm Hol}_\phi(A)$ and ${\rm Hol}_\phi(\bar A)$   are
\bea\label{ab}
\Theta_{0,A}&=&\Theta_{0,\bar A}=0~,\cr
\Theta_{1,A}&=&\Theta_{1,\bar A}=16\pi^2\CL~.
\eea

A trivial gauge transformation on $(A,\bar A)$ will leave the holonomy unchanged but will change the form of the connection, and hence the fields $g_{\mu\nu}$ and $\psi_{\mu\nu\rho}$.  In the $SL(2,\R)$ case these gauge transformations just lead to diffeomorphisms of the metric $g_{\mu\nu}$.  In the $SL(3,\R)$ case the gauge transformations will in general mix the metric $g_{\mu\nu}$ and the spin-3 field $\psi_{\mu\nu\rho}$.  Thus apparently `trivial' gauge transformations can dramatically alter the properties of the metric $g_{\mu\nu}$.   They can change the causal structure of space-time or can create a singularity from a smooth geometry.  Let us illustrate this in the following simple case.

Consider
\bea\label{ac}
A'&=b^{-1}\,a_+\,b+b^{-1}\,db~,\quad\bar A'&= b\,a_-\,b^{-1}+b\,db^{-1}
\eea
with $b=\exp(\rho L_0)$ and
\bea\label{acc}
a_+ &=&\left(L_{1}-\mathcal{L}L_{-1}+\alpha W_{-1}\right) dx^{+} \cr
a_- &=& -\left(L_{-1}-\mathcal{L}L_{1}+\bar{\alpha }W_{1}\right) dx^{-}
\eea
where $\alpha$ and $\bar \alpha$  are constants. The connections \eqref{acc} have the property that the asymptotic behavior of the space-time metric constructed from $(A',\bar A')$ is the  same as the one from $(A,\bar A)$ and that the spin-3 field does not diverge at the boundary. But more interestingly, the holonomy invariants for $(A',\bar A')$ are exactly \eqref{ab} which implies that the connections $A$ and $A'$ are thus gauge equivalent. Indeed, in appendix \ref{app:trivial} we verify that $A$ and $A'$ are related by a trivial gauge transformation as expected.

We now study the geometries associated to $(A',\bar A')$. When $\CL=-1/4$, the connection is gauge equivalent to empty AdS.  However, the metric and spin-3 field are non-trivial.  For example, if we choose $\alpha=-\bar \alpha$ we have
\bea\label{metrica}
ds'^2&=& d\rho^2-\left[(e^\rho+{1\over 4}e^{-\rho})^2-\alpha^2e^{-2\rho}\right]dt^2+\left[(e^\rho-{1\over 4}e^{-\rho})^2-\alpha^2e^{-2\rho}\right]d\phi^2~,\cr
\psi'&=&-8\alpha\, d\rho d\phi dt~.
\eea
In contrast to the usual representation of global AdS$_3$,  the
metric \eqref{metrica} is far from smooth. If $\alpha\ne 0$ the
metric becomes singular at some positive value of $\rho$ where
$g_{tt}$ and $g_{\phi\phi}$ vanish.  By performing a trivial gauge
transformation, we have created a singular space-time from a regular
one.

In fact, the reverse is true as well.  If we start with a singular metric, it is possible to remove the singularity via a $SL(3,\R)$ gauge transformation. Let's consider the simple case of the conical deficit with $-1/4<\CL<0$, the metric and spin-3 field  for $(A',\bar A')$ are
\bea
ds'^{2}=d\rho ^{2}-\left[ \left( e^{\rho }-\mathcal{L}e^{-\rho }\right)
^{2}+\alpha \bar{\alpha}e^{-2\rho }\right] dt^{2}+\left[ \left( e^{\rho }+
\mathcal{L}e^{-\rho }\right) ^{2}+\alpha \bar{\alpha}e^{-2\rho }\right]
d\phi ^{2}~,
\eea
and
\bea
\psi' =-2d\rho d\phi ^{2}\left[ 1+\mathcal{L}e^{-2\rho }\right] \left( \alpha
+\bar{\alpha}\right) -2d\rho dt^{2}\left[ 1-\mathcal{L}e^{-2\rho }\right]
\left( \alpha +\bar{\alpha}\right) +4d\rho d\phi dt\left[ \bar{\alpha}%
-\alpha \right]~.
\eea
We first note that for $\alpha\bar\alpha>0$ the components of the metric never vanish.\footnote{If $\alpha\bar\alpha<0$, the geometry contains curvature singularities.} Thus the metric is completely smooth.  Moreover, the geometry will have two boundaries since $\rho$ now ranges from plus to minus infinity. The price we pay is that the spin-3 field $\psi_{\mu\nu\alpha}$  vanishes in the interior and reaches the boundary at $\rho=-\infty$. Thus, in a sense a zero in the metric has been exchanged with a zero in the spin-3 field.

Using the enhanced gauge symmetry we were able to resolve the metric singularity of the conical deficit while keeping the physical observable (the holonomy) unchanged. While this is an appealing feature of the theory, we have not yet proven that this solution is physically admissible.  For example, it might be that the linearized excitations around this solution are stable and ghost-free.
It would be interesting to investigate this further.

\section{ $SL(3,\R)$ Black holes}

We now proceed to study in more detail the solutions of higher spin gravity which have non-trivial holonomy around a single cycle.
These will be interpreted as black holes, so before constructing the solutions we begin by describing in general the expected features of such higher spin black holes.

Our first question is how to define a black hole
in a higher spin theory, or indeed whether such solutions are
allowed. For example, in \cite{Shenker:2011zf} the authors have
conjectured that Vasiliev's theory  in AdS$_4$ should not contain
Schwarzschild type solutions. In three dimension the situation is
significantly different. The BTZ black hole is a quotient of
AdS$_3$, and hence guaranteed to be a solution to the higher spin
theory (albeit not a particularly novel one). More
interesting solutions would describe black holes which are charged
under the higher spin fields.  In \cite{Gutperle:2011kf,
Ammon:2011nk, Kraus:2011ds} such solutions were constructed. Our
goal here is to extend this analysis and present it in a more
general framework. In particular we will find that the different
embeddings of $SL(2,\RR)$ into  $SL(N,\RR)$ will lead to different
thermodynamical interpretations of the black hole.  In particular we
will see that the solutions of \cite{Gutperle:2011kf, Ammon:2011nk,
Kraus:2011ds}, when interpreted in the diagonal embedding, can be
viewed as BTZ black holes which are charged under a set of $U(1)$
Chern-Simons gauge fields.

As we saw in the previous section, any given connection may have multiple geometric interpretations. Gauge transformations do not change the holonomy properties of a connection, but  they can alter the smoothness and regularity conditions of the metric. 
Thus our first goal is to exploit this gauge symmetry to put the metric in a form which allows it to be interpreted as a black hole.

\subsection{What is a higher spin black hole? }

We begin by asking what desired properties a black hole in the higher spin theory should possess.
We will seek black hole metrics which

\begin{enumerate}
\item have a smooth BTZ limit
\item have a Lorentzian horizon and a regular Euclidean continuation
\item have a thermodynamical interpretation
\end{enumerate}
These guiding principles are the same as those proposed in
\cite{Gutperle:2011kf}.  We should also note that the implementation
of all the conditions completely agrees with  \cite{Gutperle:2011kf}
for the principal embedding.

The first condition is quite natural.  The BTZ black hole is a solution of the theory, so if the higher spin charges are independent parameters then we should be able to smoothly set them zero while keeping the mass and angular momentum finite.  These charges are in one-to-one correspondence with the trace invariants of the holonomy, so a more simple way to phrase this condition is that the holonomy of $A$ (and likewise of $\bar A$) should be controlled by two independent parameters. Conditions two and three are more subtle and will be explained below.

\subsubsection{Horizons and Regular Euclidean Geometries}\label{sec:Euc}

In general a black hole is defined as a solution with a smooth horizon in Lorentzian signature, so that (in an appropriate coordinate system) the time component of the metric has a zero at some value of the radial coordinate $\rho_h$. In Euclidean signature the geometry will be smooth at $\rho_h$ provided that Euclidean time and the angular coordinate have a definite periodicity. More explicitly, this requires
\be
(t_E,\phi)\sim (t_E,\phi)+2\pi(\beta,\beta\Omega)~,
\ee
where $\beta$ and $\Omega$ are the temperature and angular potential. A convenient way to package these periodicities is via
\be\label{thermal}
z=\phi+it_E ~,\quad z\sim z+2\pi \tau~,
\ee
with $\tau= \beta\Omega+i\beta$.

Statements  which involve the metric $g_{\mu\nu}$ are not necessarily gauge
invariant. However, the condition of smoothness on
the Euclidean solution can be translated into a gauge invariant condition on the holonomy.  The Euclidean
geometry will have two cycles: the
spatial cycle $\phi\sim\phi+2\pi$ that measures the size of
the horizon and the thermal cycle $z\sim z+2\pi\tau$. The physical charges are defined by the holonomy along
the non-contractible spatial cycle.
But the requirement of a smooth Euclidean solution implies that
the holonomy around the thermal cycle is trivial; as the contractible cycle shrinks
to zero size the connection should be single valued. More explicitly, the holonomies
\be\label{Hthermal}
{\rm Hol}_\tau(A)={\cal P}\exp(\oint_{\tau} A_z\,dz)~,\quad {\rm Hol}_\tau(\bar A)={\cal P}\exp(\oint_{\bar \tau} \bar A_{\bar z}\,d\bar z)~,
\ee
must be equal to the identity element of the gauge group.
These conditions can be viewed as equations which define $\tau$ and  constrain $(A,\bar A)$ .
We emphasize that our guiding principle here -- which is the same as that of \cite{Gutperle:2011kf} -- is that the holonomy is the only gauge invariant physical observable.  In section \ref{sec:solns} we will construct higher spin black holes where  the condition on the  holonomy is compatible with the geometrical notion of a smooth horizon.

The smoothness condition on the holonomies is straightforward to implement, but we should first comment on precisely what we mean by the statement that the holonomy around a contractible cycle is equal to the identity element of the gauge group. For example, in the $SL(2,\RR)$ theory the holonomy of BTZ around the contractible cycle is
\be\label{Hads}
{\rm Hol}(A_{\rm BTZ})=\left(\begin{array}{cc}-1 & 0 \\ 0 & -1\end{array}\right)
\ee
which is not the identity element of $SL(2,\RR)$. But there is no puzzle here: minus the identity is in the center of $SL(2,\RR)$ and hence it acts trivially on the field which are built out of the connection $A$.  This $-1$  just means that half-integer spin particles will pick up a minus sign when translated around the contractible cycle.  Moving once around the thermal circle is equivalent to a rotation by $2\pi$ around the horizon $\rho_h$, under which odd-spin states pick up a minus sign.

This reflects the fact that the Chern-Simons gauge group $SL(2,\RR)\times SL(2,\RR)$ is equal to $Spin(2,2)$ rather than the
AdS${}_3$ isometry group $SO(2,2)=SL(2,\RR)\times SL(2,\RR)/\ZZ_2$, where $\ZZ_2$ is the center.  Thus when we identify the Chern-Simons theory with the gravity theory we  obtain a gravity theory whose solutions are not just manifolds, but manifolds with a spin structure.

For the $SL(N,\RR)\times SL(N,\RR)$ higher spin theories a similar
gravity structure will occur.  However, it will turn out that
whether the gravitational theory has $SO(2,2)$ structure or
$Spin(2,2)$ structure will depend on how the gravitational
$sl(2,\RR)\times sl(2,\RR)$ is embedded into the gauge group. In
particular, when we exponentiate the $sl(2,\RR)$ algebra we may or
may not obtain minus signs which reflect the fact that it acts on
spinors. Let us first consider the diagonal embedding, where
\be\label{L0}
e^{2\pi L_0}=
 \left(
\begin{array}{ccccccc}
     -1& 0&\cdots  && &  0 \\
  0 &   -1 & &&  \vdots\\
  \vdots &  &   1& &\\
   &  &   &  \ddots& &\\
0 &\cdots&  & &&1
\end{array}
\right)~.
\ee
Thus we see that the diagonal embedding corresponds to a gravitational theory with spin structure.  Indeed, in diagonal embedding the low energy excitations include spin-3/2 fields, so a spin structure is necessary to define the theory.  For the principal embedding however, $e^{2 \pi L_0}$ is exactly equal to one and the theory does not have a spin structure.  In this case we see that it is impossible to couple spinors to the theory; this is not a problem however, as all of the low energy excitations are of integer spin.

This observation is important when we implement the smoothness condition \eqref{Hthermal}, which is an equation for $\tau$ as well as for the other parameters in the connections. Hence there will be important factors of 2, for example, in the equations for $\tau$ for different gravitational embeddings.

\subsubsection{Thermodynamics}\label{sec:thermo}

The black holes of higher spin gravity should obey the laws of thermodynamics. In the thermodynamic description of black hole solutions we should first clarify which quantities in the solution should be identified with extensive variables (potentials) versus intensive variables (charges). The solutions will be characterized by 4 independent global charges, which we denote as $(L_0,W_0)$ and $(\bar L_0,\bar W_0)$.  These charges will be constructed explicitly below; they are defined as the generators of the corresponding large gauge transformations in the bulk.  These solutions will then contribute to a partition function of the form
\be\label{Z}
Z(\tau,\bar\tau,\alpha,\bar\alpha)={\rm Tr}\left(q^{L_0}\bar q^{\bar L_0} u^{W_0}\bar u^{\bar W_0}\right)~,
\ee
where $q=\exp(2\pi i\tau)$ and $u=\exp(2\pi i\alpha)$. Here  $\tau$ and $\alpha$ are the (complex) potentials conjugate to the charges.

The physical spectrum  traced over in (\ref{Z}) -- as well as the definitions of the operators $L_0$ and $W_0$ -- will depend on the choice of embedding of $sl(2,\RR)$ in $sl(3,\RR)$. For example, for the principal  embedding the states organize into representations of the $\CW_3$ algebra \cite{Campoleoni:2010zq}. In this algebra $L_0$ and $W_0$ are operators of weight $(2,0)$ and $(3,0)$ respectively.  For the diagonal embedding  states are organized into $\CW_3^{(2)}$ representations \cite{Ammon:2011nk,Campoleoni:2011hg}. Here $L_0$ is the zero mode of stress tensor, but  $W_0$  is now the zero mode of a $(1,0)$ primary field.

In principle the partition function \eqref{Z} could be computed in the classical limit from the Euclidean action of the corresponding black hole solution.  Unfortunately, this direct calculation is difficult as the action must be  regulated using some (as yet unspecified) subtraction procedure in the presence of higher spin fields.  Nevertheless, the simple existence of this partition function provides a non-trivial constraint on the solutions, and we will see that the partition function can be computed indirectly. It is convenient to work in terms of the free energy  $F=-T\log(Z)$.  We will assume that the free energy is a differentiable function of the charges and potentials (i.e. that there is no phase transition) so that we have
\be
Q_{L_0}=-\left ({\partial F\over \partial \tau}\right)_{\alpha}~,\quad  Q_{W_0}=-\left({\partial F\over \partial \alpha}\right)_{\tau}~.
\ee
Thus
\be\label{integrab}
\left({\partial Q_{L_0} \over \partial \alpha}\right)_{\tau}=\left({\partial Q_{W_0}\over \partial \tau}\right)_{\alpha}~.
\ee
This is the same integrability condition imposed in \cite{Gutperle:2011kf}. Equation \eqref{integrab} provides a consistency check on our solutions.

Let us now compute the charges and potentials explicitly. This is a very simple task for solutions which are written in the highest weight gauge, where sources and charges are explicitly decoupled in the Chern-Simons connection.  In particular, one can read off the charges by finding global symmetries under which the connection is invariant and using standard techniques in Chern-Simons theory.  A full discussion of this prescription can be found in \cite{Banados:1994tn,Banados:1998gg,Banados:1998ta}, and its implementation in higher spin theories in \cite{Henneaux:2010xg,Campoleoni:2010zq,Gaberdiel:2011wb,Campoleoni:2011hg}. For example, in the principal embedding of $SL(3,\RR)$ the highest weight gauge (including sources) is
\bea\label{HWP}
A&=& b^{-1} a\, b + b^{-1} d\,b~,\cr
a &=&\left( L_{1}-\ell_0\, L_{-1}-w_0\,W_{-2}\right) dx^{+} +\mu \left(W_2+8w_0\,L_{-1}+ \ell_0^2\, W_{-2}-2\ell_0\,W_0 \right) dx^{-}~,
\eea
and a similar expression for $\bar A$. In this gauge we have\footnote{The normalization of charges follow from the conventions in  \cite{Ammon:2011nk,Campoleoni:2011hg} for both embeddings.}
\be
Q_{L_0}={c\over 6}\ell_0 ~,\quad Q_{W_0}={2c\over 3}w_0~,
\ee
where we have introduced the central charge
\be
c=6 k \epsilon = {3\ell\over 2G}~,
\ee
and $\epsilon$ was defined in \eqref{sec2:ba}; for the principal embedding $\epsilon=4$ and for the diagonal embedding $\epsilon=1$.
The chemical potential (source) associated to $Q_{W_0}$ is  $\alpha=\bar\tau\mu$.   Both $\tau$ and $\alpha$ will be completely fixed as a function of the charges  by the smoothness condition \eqref{Hthermal}. After imposing this condition, $Q_{L_0}$ and $Q_{W_0}$ are the two independent parameters that specify uniquely the connection.

For the diagonal embedding we have
\bea\label{HWD}
A&=& b^{-1} a\, b + b^{-1} d\,b~,\cr
a &=&\left( W_{2}+w\, W_{-2}-q\,W_{0}\right) dx^{+} +{\eta\over 2} W_{0}  dx^{-}~.
\eea
In this case
\be\label{chargediag}
Q_{L_0}={ c\over 6}\left(16 w +{4\over 3}q^2\right)~,\quad Q_{W_0}=- {2 c\over 9}{q}~.
\ee
$\alpha=\tau\eta$ is the chemical potential (source) associated to $W_0$, and as before \eqref{Hthermal}  determines $\tau$ and $\alpha$  as a function of the charges.

The advantage of the highest weight gauge is that the charges and potentials are easy to compute. However, the metric may not take a nice form in this gauge. For example, in the first attempts to construct black holes by \cite{Gutperle:2011kf,Ammon:2011nk} it was noticed that even though \eqref{HWP} satisfies all 3 conditions, the metric is not that of a black hole, i.e. there is no event horizon, and instead the solution is a  wormhole. A complicated gauge transformation takes \eqref{HWP} to a form where it is evident that the solution is a black hole.  For the solutions we will construct below we will face a similar challenge. The connections that give a smooth black hole metric do not have the form \eqref{HWP}, and constructing the gauge transformation that brings them to the highest weight gauge  is complicated.

A simple way to overcome this is to use the fact that the holonomies capture all gauge invariant information about a connection. Thus given a connection satisfying conditions 1 and 2, one can  compute the  charges and potentials by matching the holonomy with that of a connection in the highest weight gauge, either \eqref{HWP} or \eqref{HWD} depending on which part of the connection we wish to interpret as the gravitational $SL(2,\RR)$.

We now present the holonomy invariants around the non-contractible cycle for both embeddings. For the principle embedding \eqref{HWP}
\bea\label{invMP}
{\rm Tr}(a_\phi^2)&=&8\left({16\over 3} \mu^2\ell_0^2 +\ell_0+12\mu w_0\right)~,\cr
{\rm Det} (a_\phi)&=&-16 w_{0}\left(1 +16 \ell_{0}\mu^{2}\right)~,
\eea
and similar expression for $\bar a$. For the diagonal embedding  \eqref{HWD}
\bea\label{invMD}
{\rm Tr}(a_\phi^2)&=&32\left({1\over 3} q^2 +w\right)~,\cr
{\rm Det} (a_\phi)&=&{128\over 3}q\left({q^2\over 9}-w\right)~.
\eea
For both \eqref{invMP} and \eqref{invMD}, we have simplified the invariants by using the smoothness constraints on the thermal holonomy. In particular this implies that in \eqref{invMP} we should treat $\mu$ as a function of $\ell_0$ and $w_0$.

\subsection{The Solutions}\label{sec:solns}

We now construct $SL(3,\R)$ connections which describe black hole solutions with non-zero higher spin charge.  We consider the class of connections defined in \eqref{CSsol} with $b=e^{\rho L_0}$ and
\be\label{aans}
a=a_+ dx^+ +a_-dx^-~,\quad \bar a=\bar a_+ dx^+ +\bar a_-dx^-~,
\ee
where $x^\pm=t\pm \phi$. Here $a_{\pm}$ and the barred counterparts are constant $sl(3,\RR)$ matrices.
The connections are
 \bea\label{aansx}
 A=b^{-1}a\,b +L_0 d\rho~, \quad \bar A=b\,\bar a \, b^{-1}-L_0d\rho~.
 \eea
We recall that the holonomy of an $SL(3,\RR)$ connection is parameterized by two gauge-invariant pieces of data.  Thus our solutions will be labelled by four global charges: two charges each from the holonomies of the connections $A$ and $\bar A$. Two of these charges will be related to mass and angular momentum and two to non-trivial higher spin charges.

The corresponding geometries will be axisymmetric, just like the (rotating) BTZ black hole solutions.  However to keep the equations simple, we restrict attention to non-rotating solutions for which ${\rm Hol}(A)\equiv {\rm Hol}(\bar A)$. The black hole will carry mass and one higher spin charge. The generalization to 4 independent charges is completely straightforward.

\subsubsection{The simplest solution}

In the diagonal embedding of $SL(2,\RR)$ in $SL(3,\RR)$ it is particularly simple to construct black holes.  We recall that a consistent truncation of this theory is just general relativity coupled to a pair of $U(1)$ Chern-Simons gauge fields.  So we may consider the BTZ black which carries charge under these gauge fields. The connection is \eqref{aansx} with
\bea\label{3:a}
a &= &\left[W_{2}+w W_{-2}-qW_{0}\right]dx^+ +\frac{\eta}{ 2} W_0 dx^- ~,\cr
\bar a &=& \left[W_{-2}+ w W_{2}-qW_{0}\right]dx^- +\frac{\eta}{ 2} W_0 dx^+ ~.
\eea
 The metric, as defined in \eqref{sec2:da}, is
\bea\label{gbh}
ds^2&=&d\rho^2-4\left(
e^{2\rho }-w e^{-2\rho }\right) ^{2}dt^2+4\left(
e^{2\rho }+w e^{-2\rho }\right) ^{2} d\phi^2 ~,
\eea
and the pair of $U(1)$ fields are
\be
\chi= -q dx^+ +{\eta\over 2} dx^-~,\quad\bar \chi= -q dx^- +{\eta\over 2} dx^+~.
\ee
This is a standard black hole with mass $w$ and charged under both of the $U(1)$ gauge fields.

We must impose the smoothness condition and verify that the black hole obeys the first law of thermodynamics. The holonomies around the thermal cycle are
\be
\exp\left(\int_0^{2\pi\beta}A_{t_E}dt_E\right)~,\quad \exp\left(\int_0^{2\pi\beta}\bar A_{t_E}dt_E\right)~.
\ee
For the connection to be single valued around $t_E\sim t_E+2\pi \beta$ in the diagonal embedding we must have
\be\label{potD}
\eta=2q ~,\quad \beta={1\over 8\sqrt{w} }~.
\ee
This will lead to a smooth geometry, since the time component of  the $U(1)$ fields vanish at the horizon (and hence everywhere), and the periodicity of $t=it_E$  makes the euclidean continuation of \eqref{3:a} smooth.

The conserved charges can be easily determined  from \eqref{chargediag}  as a function of the potentials \eqref{potD}
\be\label{CD}
Q_{L_0}={ c\over 6}\left(-{1\over 4\tau^2}+{\alpha^2\over 3 \tau^2}\right)~,\quad Q_{W_0}=-{c\over 9}{\alpha\over \tau}~,
\ee
with $\tau=i\beta$ and $\alpha=\tau \eta$. From here it is evident that the integrability condition \eqref{integrab} is satisfied. Using
\be
\tau={1\over 2\pi i}{\partial S\over \partial Q_{L_0}}~,\quad\alpha=-{1\over 2\pi i}{\partial S\over \partial Q_{W_0}}~,
\ee
the entropy is given by
\be\label{xa}
S=2\pi \sqrt{{c\over 6} \left(Q_{L_0}-{9\over 2 c }(Q_{W_0})^{2}\right)}~.
\ee

There are two interesting features of the entropy. First, the dependence of the entropy on the charges is exactly what it is expected from a $R$-charged  BTZ black hole.\footnote{The near horizon of a BMPV black hole is an example of  a charged BTZ black hole, where the $R$ charge is identified with the five dimensional angular momentum.}   This is not a surprise, since in the language of the dual CFT the charge  $Q_{W_0}$ corresponds to a $(1,0)$ current and the combination in \eqref{xa} is the spectral flow invariant.

The second property is that \eqref{xa} is exactly the Bekenstein-Hawking entropy
\be
S= {A\over 4G}= {2\pi \ell r_+ \over 4G}~, \quad r_+=4\sqrt{w}~,
\ee
in the truncation of the higher spin theory to gravity coupled to $U(1)$ Chern-Simons fields.

We note that the connections \eqref{3:a} can also be interpreted as a black hole solution in the principal embedding. The connections \eqref{3:a} are gauge equivalent to \eqref{HWP} provided that the holonomies around the non-contractible circle are the same.
Using \eqref{invMP} and \eqref{invMD} we find that\footnote{As noted in \eqref{invMP}, $\mu$ is a function of $\ell_0$ and $w_0$. In the highest weight gauge there is a constraint on $\mu$ from requiring that the holonomy is trivial around the thermal cycle. }
\bea
-w_{0}\left(1 +16 \ell_{0}\mu^{2}\right)={8q\over 3}\left({q^2\over 9}-w\right)\cr
{16\over 3} \mu^2\ell_0^2 +\ell_0+12\mu w_0= {4\over 3} q^2 +4w
\eea
The metric and spin-3 field are
\be\label{sec5:ag}
ds^2=d\rho^2-4\left(
e^{2\rho }-w e^{-2\rho }\right) ^{2}dt^2+4\left[\left(
e^{2\rho }+w e^{-2\rho }\right) ^{2}+{1\over 3}(q+{\eta\over 2})^{2}\right] d\phi^2 ~,
\ee
and
\be
\psi=-{2\over 3}(Q+{\eta\over 2})\left[d\rho^2-4\left(
e^{2\rho }-w e^{-2\rho }\right) ^{2}dt^2+4\left(\left(
e^{2\rho }+w e^{-2\rho }\right) ^{2}-{1\over 3}(q+{\eta\over 2})^{2}\right) d\phi^2\right]d\phi~.
\ee
The smoothness condition \eqref{Hthermal} gives
\be
\eta=2q ~,\quad \beta={1\over 4\sqrt{w} }~.
\ee

The thermodynamics for the principal embedding in the highest weight gauge were carried out in \cite{Gutperle:2011kf} and we will not repeat here. The resulting entropy is given by
\be
S={2\pi\over 3} c \sqrt{\ell_0} f(C)~,
\ee
with
\be
f(C)= \sqrt{1-{3\over 4C}}~,\quad {C-1\over C^{3/2}}=\sqrt{{3\pi\over c}{w_0\over \ell_0}}~.
\ee

\subsubsection{A spin-3 black hole}

We now consider a generalization of the previous ansatz, which will give solutions satisfying all three of our conditions.  We will generalize \eqref{3:a} to
\bea\label{3:b}
a &=&\left[\ell_D W_{2}+ \CW W_{-2}-QW_{0}\right]dx^+ +\left[\ell_P L_{1}-\CL L_{-1}+{\Phi} W_0\right] dx^-~,\cr
\bar a &=&\left[\ell_D W_{-2}+ \CW W_{2}-QW_{0}\right]dx^- -\left[\ell_P L_{-1}-\CL L_{1}-{\Phi} W_0\right] dx^+~.
\eea
The equations of motion imply
\be
Q={2 \CW \ell_P\over \CL}~,\quad {\CL^2\over \ell_P^2}= { \CW \over \ell_D} ~.
\ee

\noindent In the principal embedding, the metric  for this configuration reads
\bea\label{gbh}
ds^2&=&d\rho^2-\left[4\ell_D^2\left(e^{2\rho }-{\CL^2 \over \ell_P^2}e^{-2\rho }\right)^{2}+\ell_P^2\left( e^{\rho }-{\CL \over \ell_P} e^{-\rho }\right)^{2}\right]dt^2\cr
&&+\left[4 \ell_D^2\left(e^{2\rho }+{\CL^2 \over \ell_P^2}e^{-2\rho }\right)^{2}+\ell_P^2\left( e^{\rho }+{\CL \over \ell_P} e^{-\rho }\right)^{2}+{4\over 3}(Q+\Phi)^2\right]d\phi^2~.
\eea
The trivial holonomy constraint give
\be\label{temp}
\beta^2={\CL^2\over 4\ell_P}(16\CW^2\ell_P+\CL^3)^{-1}~,\quad \Phi=8{\CW \ell_P\over \CL}~.
\ee
We note that the periodicity of Euclidean time that assures regularity of the metric \eqref{gbh} is precisely given by the period in  \eqref{temp}, that is
\be
\beta = \sqrt{\frac{-2}{g_{tt}^{\prime \prime }|_{h}}}~.
\ee
Furthermore, the $\psi_{\phi t t}$ component of the spin-3 field vanishes at the horizon and the periodicity as defined in \cite{Gutperle:2011kf} at this point matches the one obtained from the metric\footnote{In appendix \ref{app:spin} we give the exact expression for $\psi_{\mu\nu\lambda}$ that supports \eqref{gbh}.}
\be
\beta = \sqrt{ \frac{-2 \psi_{\phi \rho \rho}}{\psi_{\phi t t}^{\prime \prime }} \Big\vert_{h}}~.
\ee
The appeal of this solution is that it gives a rather simple black hole in the principal embedding which has all the geometrical properties we would expect. From the structure of the line element \eqref{gbh} it  is clear that  the BTZ can be obtained smoothly for either the limit $\ell_D=\CW=0$ or $\ell_P=\CL=0$.  

We emphasize that the solution \eqref{3:b} is gauge equivalent to \eqref{3:a}, so we haven't really found a new solution. We simply performed a gauge transformation where the black hole has a horizon in the principal embedding rather than the diagonal embedding. By matching the holonomies of both solutions the map between the parameters in \eqref{3:a} and \eqref{3:b} is
\be
q=-4{\CW \ell_P\over \CL}~, \quad w={\ell_P\over 4\CL^2}(16 \CW^2\ell_P+\CL^3)
\ee
The remaining analysis of thermodynamics is then equivalent to the discussion in the previous section.

\section{Discussion}

We have successfully defined and constructed higher spin black holes in agreement with previous work \cite{Gutperle:2011kf}. The metric-like fields and thermodynamics of the solution depend on the embedding, i.e. how we interpret the Chern-Simons gauge theory as a gravitational theory. What we have exploited here is that the theory can be truncated to gravity coupled to $U(1)$ Chern-Simons fields, where it is almost trivial to construct a black hole that meets our geometric expectations as well as satisfying the gauge invariant constraints on the holonomy. We also used this simplicity to generalize the ansatz so that the black hole has a natural interpretation in terms of the metric-like fields for the principal embedding.

The solutions constructed here are non-extremal black holes, i.e. the mass is not a function of the charges and the temperature is finite. This is reflected in the fact that the two eigenvalues for the holonomy around the non-contractible cycle are independent. Of course there will be extremal solutions in the theory, and in analogy to the extremal BTZ black hole, those solutions would correspond to connections whose holonomies around $\phi$ are not diagonalizable. It would be interesting to explore if this class of extremal solutions --i.e.  connections with  non-diagonal Jordan decomposition-- requires that the black hole is at zero temperature.


%
%

\section*{Acknowledgements}

We are grateful to Max Ba\~nados, Andrea Campoleoni, Rajesh Gopakumar, Michael Gutperle, Tom Hartman,  Per Kraus, Josh Lapan and
Finn Larsen for very useful discussions.  This work was supported by
the National Science and Engineering Research Council of Canada and
FQRNT (Fonds qu\'eb\'ecois de la recherche sur la nature et les
technologies). E.H. acknowledges support from Fundaci\'on Caja
Madrid.

\appendix

\section{Conventions}\label{app:conv}

 We will work in a representation such that the matrices obey \eqref{algebra}, and the Lie algebra metric is
\begin{eqnarray}\label{liemetric}
\mathrm{tr}(J_{a}J_{b}) &=&2\eta _{ab}~, \cr
\mathrm{tr}(J_{a}T_{bc}) &=&0~, \cr
\mathrm{tr}(T_{ab}T_{cd}) &=&-\frac{4}{3}\eta _{ab}\eta _{cd}+2\left( \eta
_{ac}\eta _{bd}+\eta _{ad}\eta _{bc}\right)~.
\end{eqnarray}

A more convenient basis for $J_a$ and $T_{ab}$ is given by the generators $L_i$ and $W_m$ which satisfy
\bea
[L_i ,L_j] &=& (i-j)L_{i+j}~, \cr
[L_i, W_m] &=& (2i-m)W_{i+m}~, \cr
[W_m,W_n] &=& -{1 \over 3}(m-n)(2m^2+2n^2-mn-8)L_{m+n}~.
\eea

The generators are related via the isomorphism
\be\label{app:JL}
J_0={1\over 2}(L_1+L_{-1})~,\quad J_1={1\over 2}(L_1-L_{-1})~, \quad J_2=L_0~,
\ee
and
\bea\label{app:TW}
T_{00}&=&{1\over 4}(W_2+W_{-2}+2W_0)~,\quad T_{01}={1\over 4}(W_2-W_{-2})~,\nonumber \\
T_{11}&=&{1\over 4}(W_2+W_{-2}-2W_0)~,\quad T_{02}={1\over 2}(W_1+W_{-1})~,\nonumber\\
T_{22}&=&W_0~,\quad \quad \quad\quad \quad \quad\quad \quad~~ T_{12}={1\over 2}(W_1-W_{-1})~.
\eea
The Lie algebra metric in this basis is
\bea
{\rm tr} ( L_0 L_0) &=& 2~,\quad   {\rm tr} ( L_1 L_{-1} ) = -4 ~,\cr   {\rm tr} ( W_0 W_0 )  &= &{8 \over 3}~,\quad {\rm tr} ( W_1 W_{-1} )  = -4 ~,\quad {\rm tr} ( W_2 W_{-2} )  = 16~.
\eea
and the explicit representation of the matrices is
\bea
L_1 & =& \bmat 0&0&0 \\ 1&0&0 \\ 0&1&0\emat,\quad L_0= \bmat 1&0&0 \\ 0&0&0 \\ 0&0&-1\emat,\quad  L_{-1} = \bmat 0&-2&0 \\ 0&0&-2 \\ 0&0&0\emat~,\cr &&\cr
W_2 &= &2 \bmat 0&0&0 \\ 0&0&0 \\ 1&0&0\emat,\quad W_1 =  \bmat 0&0&0 \\ 1&0&0 \\ 0&-1&0 \emat,\quad  W_0 = {2\over 3}\bmat 1&0&0 \\ 0&-2&0 \\ 0&0&1\emat~, \cr & &\cr
W_{-1} &= &\bmat 0&-2&0 \\ 0&0&2 \\ 0&0&0\emat,\quad W_{-2} = 2\bmat 0&0&4 \\ 0&0&0 \\ 0&0&0\emat~.
\eea

\section{Trivial gauge transformation for AdS$_3$}\label{app:trivial}

Looking back at our connections  $A$ in \eqref{aa} and $A'$ in \eqref{ac}. We want to write $A'$ as
\bea
A' &=&g_{\rm new}^{-1}dg_{\rm new}\cr
&=&\left( g_{\Lambda}^{-1}g_{\rm AdS}^{-1}\right) \left(
dg_{\rm AdS}\,g_{\Lambda }+g_{\rm AdS}\,dg_{\Lambda }\right) \cr
&=&g_{\Lambda }^{-1}A \,g_{\Lambda }+g_{\Lambda }^{-1}dg_{\Lambda }
\eea
 where
\be
g_{\rm new}=g_{\rm AdS}g_{\Lambda}~,
\ee
and
\be
A=g_{\rm AdS}^{-1}dg_{\rm AdS}~.
\ee
The group elements $ g_{\rm AdS}$  and $g_{\rm new}$ are known from \eqref{aa} and \eqref{ac}, so it is straight forward to solve for  $g_{\Lambda}$
\bea
g_{\Lambda } &=&g_{\rm AdS}^{-1}g_{\rm new}\cr
&=&e^{-\rho L_{0}}e^{-x^{+}\left( L_{1}+
\frac{1}{4}L_{-1}\right) }e^{x^{+}\left( L_{1}+\frac{1}{4}L_{-1}+\alpha
W_{-1}\right) }e^{\rho L_{0}}
\eea
By exponentiating the matrix $g_{\Lambda}$ explicitly, one can see that it is invariant under the identification $\phi\sim \phi+2\pi $. We conclude that $g_{\Lambda}$ is a trivial gauge transformation.

\section{Spin-3 field}\label{app:spin}
For completeness, here we record the spin-3 field supported by the connections \eqref{3:b}
\bea
\psi_{\phi\rho\rho}&=&-{20\over 3}{\CW \ell_{P}\over \CL}~,\cr
\psi_{\phi tt}&=&{20\over 3}{\CW \ell_{P}\over \CL}\left[4\ell_D^2\left(e^{2\rho }-{\CL^2 \over \ell_P^2}e^{-2\rho }\right)^{2}+\ell_P^2\left( e^{\rho }-{\CL \over \ell_P} e^{-\rho }\right)^{2}\right] +{3\CW \ell_P^{4} \over \CL^{2}}\left( e^{\rho }-{\CL \over \ell_P} e^{-\rho }\right)^{4}~,\cr
\psi_{\phi\phi\phi}&=&-{20\over 3}{\CW \ell_{P}\over \CL}\left[4\ell_D^2\left(e^{2\rho }+{\CL^2 \over \ell_P^2}e^{-2\rho }\right)^{2}+\ell_P^2\left( e^{\rho }+{\CL \over \ell_P} e^{-\rho }\right)^{2}\right] +{\CW \ell_P^{4} \over \CL^{2}}\left( e^{\rho }+{\CL \over \ell_P} e^{-\rho }\right)^{4}\cr
& & +{8\CW \ell_P^{3} \over \CL}\left( e^{\rho }+{\CL \over \ell_P} e^{-\rho }\right)^{2} +\left({20 \over 3}{\CW\ell_P\over \CL}\right)^{3}~.
\eea

\bibliographystyle{utphys}
\bibliography{all}

\end{document}